\documentclass[nofootinbib,aps,pra,showpacs,superscriptaddress,preprint]{revtex4}

\usepackage{amsmath}

\usepackage{subfig}

\usepackage{graphicx}

\usepackage{lxfonts}

\catcode`ð=\active

\defð{\u{g}}

\catcode`Ð=\active

\defÐ{\u{G}}

\catcode`Ý=\active

\defÝ{\. I}

\catcode`ö=\active

\defö{\"{o}}

\catcode`Ö=\active

\defÖ{\"O}

\catcode`ü=\active

\defü{\"{u}}

\catcode`Ü=\active

\defÜ{\"{U}}

\catcode`Þ=\active

\defÞ{\c{S}}

\catcode`þ=\active

\defþ{\c{s}}

\catcode`ý=\active

\defý{{\i}}

\catcode`ç=\active

\defç{\d{c}}

\catcode`Ç=\active

\defÇ{\d{C}}

\begin{document}

\title{Exact Solutions of the Schrödinger Equation via Laplace Transform Approach:
Pseudoharmonic potential and Mie-type potentials}
\author{\small Altuð Arda}
\email[E-mail: ]{arda@hacettepe.edu.tr}\affiliation{Department of
Physics Education, Hacettepe University, 06800, Ankara,Turkey}
\author{\small Ramazan Sever}
\email[E-mail: ]{sever@metu.edu.tr}\affiliation{Department of
Physics, Middle East Technical  University, 06531, Ankara,Turkey}

\begin{abstract}
Exact bound state solutions and corresponding normalized
eigenfunctions of the radial Schrödinger equation are studied for
the pseudoharmonic and Mie-type potentials by using the Laplace
transform approach. The analytical results are obtained and seen
that they are the same with the ones obtained before. The energy
eigenvalues of the inverse square plus square potential and
three-dimensional harmonic oscillator are given as special cases.
It is shown the variation of the first six normalized
wavefunctions of the above potentials. It is also
given numerical results for the bound states of two diatomic molecular potentials, and compared the results with the ones obtained in literature.
\end{abstract}

\maketitle

\newpage

\section{Introduction}
Molecular vibrational and rotational spectroscopy is one of the
important parts of molecular physics and one of the main tools for
other scientific areas such as biology [1] and environmental
sciences [2]. The harmonic oscillator could be useful ground to
explain the molecular vibrations but this model is restricted for
only lowest states [3]. To improve the theory of molecular
vibrations the anharmonic oscillators such as Morse and Mie-type
potentials (Kratzer potential and its generalization) can be used
to solve exactly the Schrödinger equation (SE) and provide more
reliable model for diatomic molecules [4]. Mie-type potentials
being considered as an example in the present work have important
advantages such as having eigenfunctions behaving correctly at $r
\rightarrow 0$ and $r \rightarrow \infty$ and providing exact
solutions to the SE [5]. So, these potentials have been used to
determine molecular structures and received much attention in
literature [6]. In the present work, we deal with also another
diatomic potential called pseudoharmonic potential proposed by
Davidson [7]. This potential is used to describe the
roto-vibrational states of diatomic molecules and nuclear
rotations and vibrations [8].

In the light of the above considerations, it could be interesting
to solve exactly the SE for the pseudoharmonic potential and
Mie-type potentials and find any $\ell$-state solutions in the
view of molecular physics phenomenon. Moreover, obtaining the
exact solutions of the SE for the molecular potentials is one of
the main problems in quantum physics [6]. One of the methods
giving exact solutions of the SE is used in Ref. [9] where the
wave equation is solved for the non-central potential within the
framework of the supersymmetric quantum mechanics. In Ref. [10],
the energy eigenvalues of the radial SE are obtained for the
Coulomb potential by using path integral formalism and the author
also stated how can be obtained the wave functions. In Ref. [11],
energy spectrum of the Coulomb, Morse and harmonic oscillator
potentials have been studied by using point canonical
transformation where the formalism has been extended to the case
of position-dependent mass.

We list some methods used in literature to solve the wave
equations for the pseudoharmonic potential and Mie-type
potentials: Nikiforov-Uvarov method [12, 13], algebraic approach
[14], polynomial solution [15], exact quantization rule [16],
hypervirial theorem with perturbation theory [17],
shape-invariance procedure [18], solutions in terms of
hypergeometric functions [19], \textit{etc.}. In this work, we
find exact bound state solutions of pseudoharmonic potential and
Mie-type potentials by reducing the SE to a first-order
differential equation via Laplace transform approach (LTA) and
therefore we make use of integral to the obtain energy eigenvalues
and the corresponding eigenfunctions. Actually, the LTA is an
integral transform which has been used by many authors to solve
the SE for different potentials [20, 21, 22, 23]. The LTA could be
a nearly new formalism in the literature and serve as a powerful
algebraic treatment for solving the second-order differential
equations. As a result, the LT methods describe a simple way for
solving of radial and one-dimensional differential equations. The
other advantage of this approach is that a second-order equation
can be converted into more simpler form whose solutions may be
obtained easily [22].

\section{Energy Eigenvalue Solutions}
Time-independent Schrödinger equation is written as
\begin{eqnarray}
\left\{-\frac{\hbar^2}{2m}\nabla^2+V(r)\right\}\Psi(r,\theta,\varphi)=E_{n\ell}\Psi(r,\theta,\varphi)\,,
\end{eqnarray}
and defining the wave function
$\Psi(r,\theta,\varphi)=\frac{1}{r}R(r)Y(\theta,\varphi)$, we
obtain the radial SE as [6]
\begin{eqnarray}
\left\{\frac{d^2}{dr^2}-\frac{\ell(\ell+1)}{r^2}+\frac{2m}{\hbar^2}\left[E_{n\ell}-V(r)\right]\right\}R(r)=0\,.
\end{eqnarray}
where $\ell$ is the angular momentum quantum number, $m$ is the
particle mass moving in the potential field $V(r)$ and $E_{n\ell}$
is the nonrelativistic energy of particle.

\subsection{Pseudoharmonic Potential}
The pseudoharmonic potential is given [8]
\begin{eqnarray}
V(r)=a_{1}r^2+\frac{a_2}{r^2}+a_{3}\,,
\end{eqnarray}
where $a_{i} (i=1, 2, 3)$ are real parameters. Inserting Eq. (3)
into Eq. (2), we obtain
\begin{eqnarray}
\left\{\frac{d^2}{dr^2}-\mu^2r^2-\frac{\nu(\nu+1)}{r^2}+\varepsilon^2\right\}R(r)=0\,,
\end{eqnarray}
where
\begin{eqnarray}
\mu^2=\frac{2ma_{1}}{\hbar^2}\,\,;\nu(\nu+1)=\frac{2ma_{2}}{\hbar^2}
+\ell(\ell+1)\,\,;\varepsilon^2=\frac{2m}{\hbar^2}\,(E_{n\ell}-a_{3})\,.
\end{eqnarray}
Defining the new variable $y=r^2$ and rewriting the radial wave
function as $R(y)=y^{-\nu/2}\phi(y)$, Eq. (4) turns into
\begin{eqnarray}
\left\{y\frac{d^2}{dy^2}-\left(\nu-\frac{1}{2}\right)\frac{d}{dy}
-\frac{1}{4}\left(\mu^2y-\varepsilon^2\right)\right\}\phi(y)=0\,,
\end{eqnarray}

By using the Laplace transform defined as [24]
\begin{eqnarray}
\mathcal{L}\left\{\phi(y)\right\}=f(t)=\int_{0}^{\infty}dy
e^{-ty}\phi(y)\,,
\end{eqnarray}
Eq. (6) reads
\begin{eqnarray}
\left(t^2-\frac{\mu^2}{4}\right)\frac{df(t)}{dt}+\left\{\left(\nu+\frac{3}{2}\right)t
-\frac{\varepsilon^2}{4}\right\}f(t)=0\,,
\end{eqnarray}
which is a first-order ordinary differential equation and its
solution is simply given
\begin{eqnarray}
f(t)=N\left(t+\frac{\mu}{2}\right)^{-\frac{\varepsilon^2}{4\mu}-\frac{1}{2}\left(\nu+\frac{3}{2}\right)}
\left(t-\frac{\mu}{2}\right)^{\frac{\varepsilon^2}{4\mu}-\frac{1}{2}\left(\nu+\frac{3}{2}\right)}\,,
\end{eqnarray}
where $N$ is a integral constant. In order to obtain finite wave
functions, it should be
\begin{eqnarray}
\frac{\varepsilon^2}{4\mu}-\frac{1}{2}\left(\nu+\frac{3}{2}\right)=n\,,\,\,\,(n=0,
1, 2, 3, \ldots)
\end{eqnarray}
which gives single-valued wave functions. By using this
requirement and expanding Eq. (9) into series, we get
\begin{eqnarray}
f(t)=N'\sum_{k=0}^{n}\frac{(-1)^{k}n!\left(t+\frac{\mu}{2}\right)^{-\left(\nu+\frac{3}{2}+k\right)}}{(n-k)!k!}\,,
\end{eqnarray}
where $N'$ is a constant. By using the inverse Laplace
transformation [24] we immediately obtain the solution of Eq. (6)
\begin{eqnarray}
\phi(y)=N''\sum_{k=0}^{n}\frac{(-1)^{k}n!}{(n-k)!k!}
\frac{\Gamma(\nu+\frac{3}{2})}{\Gamma(\nu+\frac{3}{2}+k)}y^{\left(\nu+\frac{1}{2}+k\right)}e^{-\mu
y/2}\,,
\end{eqnarray}
where $N''$ is a constant. On the other hand, the confluent
hypergeometric functions is defined as a series expansion [25]
\begin{eqnarray}
_{1}F_{1}(-n,\sigma,z)=\sum_{m=0}^{n}\frac{(-1)^{m}n!}{(n-m)!m!}\frac{\Gamma(\sigma)}{\Gamma(\sigma+m)}y^{m}\,,
\end{eqnarray}
So, comparing Eq. (12) with Eq. (13) we deduce that
\begin{eqnarray}
\phi(y)=N'''e^{-\mu
y/2}y^{\nu+\frac{1}{2}}\,_{1}F_{1}(-n,\nu+\frac{3}{2},y)\,,
\end{eqnarray}
We obtain finally the radial wave functions
\begin{eqnarray}
R(y)=\mathcal{N}e^{-\mu
y/2}y^{(\nu+1)/2}\,_{1}F_{1}(-n,\nu+\frac{3}{2},y)\,.
\end{eqnarray}
where $\mathcal{N}$ is normalization constant. Using the
normalization condition given as
$\int_{0}^{\infty}\left[R(r)\right]^2dr=1$ and the relation
between the Laguerre polynomials and confluent hypergeometric
functions as
$L_{n}^{p}(x)=\frac{\Gamma(n+p+1)}{n!\Gamma(p+1)}\,_{1}F_{1}(-n,p+1,x)$
[25] , the normalization constant in Eq. (15) is written
\begin{eqnarray}
\mathcal{N}=\mu^{\left(\nu+3/2\right)/2}\sqrt{\frac{2\Gamma\left(n+\nu+\frac{3}{2}\right)}{n!}\,}\left[\Gamma\left(\nu+\frac{3}{2}\right)\right]^{-1}\,,
\end{eqnarray}
where we have used [25]
\begin{eqnarray}
\int_{0}^{\infty}x^{q}e^{-x}L_{n}^{q}(x)L_{n'}^{q}(x)dx=\frac{\Gamma(q+n+1)}{n!}\,\delta_{nn'}\,.
\end{eqnarray}

In Fig. (1) we show the variation of the normalized wave functions
of the pseudoharmonic potential on the coordinate $r$. We give
first six wave functions according to the quantum number pairs
$(n,\ell)$. Inserting the parameters in Eq. (5) into Eq. (10), the
energy spectrum of the pseudoharmonic potential is obtained
\begin{eqnarray}
E_{n\ell}=a_{3}+\sqrt{\frac{8\hbar^2a_{1}}{m}\,}
\left(n+\frac{1}{2}+\frac{1}{4}\sqrt{1+4\ell(\ell+1)+\frac{8ma_{2}}{\hbar^2}\,}\right)\,.
\end{eqnarray}
We give our numerical energy eigenvalues for two different
diatomic potentials in Table I. We compare our results with the
ones given in Ref. [12] by setting the potential parameters as
$a_{1}=\frac{D_0}{r^2_{0}},\,a_{2}=D_{0}r^2_{0}$ and
$a_{3}=-2D_{0}$. Let us study the results of some special cases.
Firstly, if we put $a_{3}=0$ we obtain from Eq. (18)
\begin{eqnarray}
E_{n\ell}=\sqrt{\frac{8\hbar^2a_{1}}{m}\,}
\left(n+\frac{1}{2}+\frac{1}{4}\sqrt{1+4\ell(\ell+1)+\frac{8ma_{2}}{\hbar^2}\,}\right)\,,
\end{eqnarray}
which is the same result given in Ref. [26] for the potential of
the form $\frac{A}{r^2}+Br^2$. Secondly, if we choose the
parameters as $a_{2}=a_{3}=0$ and $a_{1}=\frac{1}{2}m\omega^2$ in
Eq. (18) we get
\begin{eqnarray}
E_{n\ell}=\hbar\omega\left(2n+\ell+\frac{3}{2}\right)\,,
\end{eqnarray}
where if we define $n'=2n+\ell$ as 'principal quantum number' we
obtain
\begin{eqnarray}
E_{n'\ell}=\hbar\omega\left(n'+\frac{3}{2}\right)\,.
\end{eqnarray}
which is exactly the spectrum of three-dimensional harmonic
oscillator [26].

\subsection{Mie-type Potentials}
The Morse potential is an example of this type of potentials or
Kratzer potential and its generalization having the forms,
respectively,
\begin{eqnarray}
V(r)=-D\left(\frac{2r_{0}}{r}-\frac{r^2_{0}}{r^2}\right)\,,
\end{eqnarray}
and
\begin{eqnarray}
V(r)=D\left(\frac{r-r_{0}}{r}\right)^2\,.
\end{eqnarray}
where $D$ is the dissociation energy and $r_{0}$ is the
equilibrium distance [15]. So, the Mie-type potentials can be
simply given as
\begin{eqnarray}
V(r)=\frac{a}{r^2}+\frac{b}{r}+c\,,
\end{eqnarray}
where $a, b, c$ are real potential parameters. Inserting Eq. (24)
into Eq. (2), redefining the wave function as
$R(r)=\sqrt{r\,}\varphi(r)$ and using the following abbreviations
\begin{eqnarray}
\gamma^2=\frac{2ma}{\hbar^2}+\ell(\ell+1)+\frac{1}{4}\,\,;
\delta^2=\frac{2mb}{\hbar^2}\,\,;\varepsilon^2=\frac{2m}{\hbar^2}(c-E_{n\ell})\,,
\end{eqnarray}
gives
\begin{eqnarray}
\left\{r^2\frac{d^2}{dr^2}+r\frac{d}{dr}-\left[\frac{\gamma^2}{r^2}+\frac{\delta^2}{r}+\varepsilon^2\right]r^2\right\}
\varphi(r)=0\,.
\end{eqnarray}
Setting $\varphi(r)=r^{\alpha}\phi(r)$ with $\alpha$ is a constant
and then inserting into Eq. (26) leads
\begin{eqnarray}
\left\{r^2\frac{d^2}{dr^2}+(2\alpha+1)r\frac{d}{dr}-\varepsilon^2r^2-\delta^2r+\alpha^2-\gamma^2\right\}\phi(r)=0\,,
\end{eqnarray}
In order to obtain a finite wave function when $r \rightarrow
\infty$, we must take $\alpha=-\gamma$ in Eq. (27)  and then we
get
\begin{eqnarray}
\left\{r\frac{d^2}{dr^2}-(2\gamma-1)\frac{d}{dr}-\delta^2-\varepsilon^2
r\right\}\phi(r)=0\,.
\end{eqnarray}
Applying the Laplace transform to Eq. (28) we obtain a first-order
differential equation
\begin{eqnarray}
\left(t^2-\varepsilon^2\right)\frac{df(t)}{dt}+\left[\left(2\gamma+1\right)t+\delta^2\right]f(t)=0\,,
\end{eqnarray}
whose solution is
\begin{eqnarray}
f(t)=N\left(t+\varepsilon\right)^{-(2\gamma+1)}
\left(\frac{t-\varepsilon}{t+\varepsilon}\right)^{-\frac{\delta^2}{2\varepsilon}\,-\,\frac{2\gamma+1}{2}}\,.
\end{eqnarray}
The wave functions must be single-valued which requiring that
\begin{eqnarray}
-\frac{\delta^2}{2\varepsilon}-\frac{2\gamma+1}{2}=n\,,\,\,\,(n=0.,
1, 2, 3, \ldots)
\end{eqnarray}
Taking into account this requirement and applying a simple series
expansion to Eq. (30) gives
\begin{eqnarray}
f(t)=N'\sum_{k=0}^{n}\frac{(-1)^{k}n!}{(n-k)!k!}\,(2\varepsilon)^{k}\left(t+\varepsilon\right)^{-(2\gamma+1)-k}\,,
\end{eqnarray}
where $N'$ is a constant. Using the inverse Laplace transformation
[24] in Eq. (32) we deduce that
\begin{eqnarray}
\phi(r)=N''r^{2\gamma}e^{-\varepsilon
r}\sum_{k=0}^{n}\frac{(-1)^{k}n!}{(n-k)!k!}\frac{\Gamma(2\gamma+1)}{\Gamma(2\gamma+1+k)}\,(2\varepsilon
r)^{k}\,,
\end{eqnarray}

Finally, we obtain
\begin{eqnarray}
\varphi(r)=N'''r^{\gamma}e^{-\varepsilon
r}\sum_{k=0}^{n}\frac{(-1)^{k}n!}{(n-k)!k!}\frac{\Gamma(2\gamma+1)}{\Gamma(2\gamma+1+k)}\,(2\varepsilon
r)^{k}\,,
\end{eqnarray}
Comparing last equation with Eq. (13) we write the radial wave
function as
\begin{eqnarray}
R(r)=\mathcal{N}r^{\gamma+\frac{1}{2}}e^{-\varepsilon
r}\,_{1}F_{1}(-n, 2\gamma+1,2\varepsilon r)\,.
\end{eqnarray}
where the normalization constant is given by following the same
procedure in previous section as
\begin{eqnarray}
\mathcal{N}=\Gamma(2\gamma+1)\sqrt{\frac{n!(2n+2\gamma+1)}{\Gamma(n+2\gamma+1)}\,}\,.
\end{eqnarray}
We give the dependence of the wave functions of the Kratzer
potential on spatially coordinate $r$ in Fig. (2) where the wave
functions are plotted for the same quantum number values as in
pseudoharmonic potential. Using Eqs. (25) and (31) we obtain the
energy eigenvalues of the Mie-type potentials
\begin{eqnarray}
E_{n\ell}=c-\frac{\hbar^2}{8m}\left[
\frac{2mb/\hbar^2}{n+\frac{1}{2}\left(1+2\sqrt{\frac{2ma}{\hbar^2}+\ell(\ell+1)+\frac{1}{4}\,}\,\right)}\right]^2\,.
\end{eqnarray}
which is the same result with the ones obtained in Ref. [27]. We
summarize our numerical results for different quantum number pairs
$(n, \ell)$ in Table II. To compare our results we chose the
potential parameters as
$a=D_{e}r^2_{e},\,b=-2D_{e}r_{e},\,c=D_{e}$ used in Ref. [13].

\section{Results}

Our numerical energy eigenvalues of two diatomic molecules
interacting in short-range [28] given in Tables I and II have a
good accuracy with the ones obtained in Refs. [15] and [13]. Figs.
1 and 2 show the variation of the wave functions versus $r$ for
the pseudoharmonic and Kratzer potentials. The wave functions of
the pseudoharmonic (Kratzer potential) go to zero as $r
\rightarrow 0$ and as $r \sim 8$ ($r \sim 35$). The eigenfunctions
corresponding to $n=1$ for each potential go to zero from lower
part of the vertical axes while the remaining functions from upper
part of the zero axes.

\section{Conclusions}

We have exactly solved the radial Schrödinger equation for the
pseudoharmonic and Mie-type potentials by using Laplace transform
approach. We have found the energy eigenvalues and the
corresponding normalized eigenfunctions of the diatomic
potentials. We discussed briefly some special cases of the
potentials. We observed that our analytical results and also the
results for the special cases are the same with the ones obtained
in literature. We also summarized our numerical energy eigenvalues
for two different diatomic molecules. It seems that the Laplace
transform approach is very economical method about solving the
wave equations for some potentials.

\section{Acknowledgments}
This research was partially supported by the Scientific and
Technical Research Council of Turkey.

\newpage

\begin{table}
\begin{ruledtabular}
\caption{Energy eigenvalues of the pseudoharmonic potential for
different values of $n$ and $\ell$ in $eV$ (the parameter values
are used in Ref. [12]: $D_{0}=$96288.03528 cm$^{-1}$,
$r_{0}=$1.0940 $\AA$, $m=$7.00335 amu for $N_{2}$;
$D_{0}=$87471.42567 cm$^{-1}$, $r_{0}=$1.1282 $\AA$, $m=$6.860586
amu for $CO$).}
\begin{tabular}{@{}ccllcc@{}}
&&\multicolumn{2}{c}{$N_{2}$} &\multicolumn{2}{c}{$CO$}
\\ \cline{3-4} \cline{5-6}
$n$ & $\ell$  & Our Results & Ref. [15] & Our Results & Ref. [15]  \\
0 & 0 & 0.109180 & 0.1091559 & 0.101953 & 0.1019306 \\
1 & 0 & 0.327414 & 0.3273430 & 0.305738 & 0.3056722 \\
  & 1 & 0.327913 & 0.3278417 & 0.306217 & 0.3061508 \\
2 & 0 & 0.545648 & 0.5455302 & 0.509524 & 0.5094137 \\
  & 1 & 0.546147 & 0.5460288 & 0.510003 & 0.5098923 \\
  & 2 & 0.547145 & 0.5470260 & 0.510961 & 0.5108495 \\
3 & 0 & 0.763883 &  & 0.713310 &  \\
  & 1 & 0.764382 &  & 0.713789 &  \\
  & 2 & 0.765380 &  & 0.714747 &  \\
  & 3 & 0.766877 &  & 0.716183 &  \\
4 & 0 & 0.982117 & 0.9819045 & 0.917095 & 0.9168969 \\
  & 1 & 0.982616 & 0.9824031 & 0.917574 & 0.9173755 \\
  & 2 & 0.983614 & 0.9834003 & 0.918532 & 0.9183327 \\
  & 3 & 0.985111 & 0.9848961 & 0.919969 & 0.9197684 \\
  & 4 & 0.987107 & 0.9868903 & 0.921885 & 0.9216825 \\
\end{tabular}
\end{ruledtabular}
\end{table}

\newpage

\begin{table}
\begin{ruledtabular}
\caption{Energy eigenvalues of the Kratzer potential for different
values of $n$ and $\ell$ in $eV$.}
\begin{tabular}{@{}ccllcc@{}}
&&\multicolumn{2}{c}{$N_{2}$} &\multicolumn{2}{c}{$CO$}
\\ \cline{3-4} \cline{5-6}
$n$ & $\ell$  & Our Results & Ref. [13] & Our Results & Ref. [13]  \\
0 & 0 & 0.054434 & 0.054430 & 0.050827 & 0.050823 \\
1 & 0 & 0.162068 & 0.162057 & 0.151296 & 0.151287 \\
  & 1 & 0.162557 & 0.162546 & 0.151765 & 0.151755 \\
2 & 0 & 0.268245 & 0.268229 & 0.250369 & 0.250354 \\
  & 1 & 0.268728 & 0.268711 & 0.250831 & 0.250816 \\
  & 2 & 0.269692 & 0.269675 & 0.251756 & 0.251744 \\
3 & 0 & 0.372992 & 0.372972 & 0.348070 & 0.348051 \\
  & 1 & 0.373468 & 0.373447 & 0.348526 & 0.348507 \\
  & 2 & 0.374419 & 0.374398 & 0.349438 & 0.349418 \\
  & 3 & 0.375846 & 0.375823 & 0.350806 & 0.350785 \\
4 & 0 & 0.476334 & 0.476313 & 0.444425 & 0.444403 \\
  & 1 & 0.476803 & 0.476779 & 0.444871 & 0.444852 \\
  & 2 & 0.477742 & 0.477717 & 0.445774 & 0.445751 \\
  & 3 & 0.479150 & 0.479124 & 0.447123 & 0.447099 \\
  & 4 & 0.481026 & 0.480999 & 0.448921 & 0.448895 \\
\end{tabular}
\end{ruledtabular}
\end{table}

\newpage

\begin{figure}
\centering
\includegraphics[height=5in, width=6.5in, angle=0]{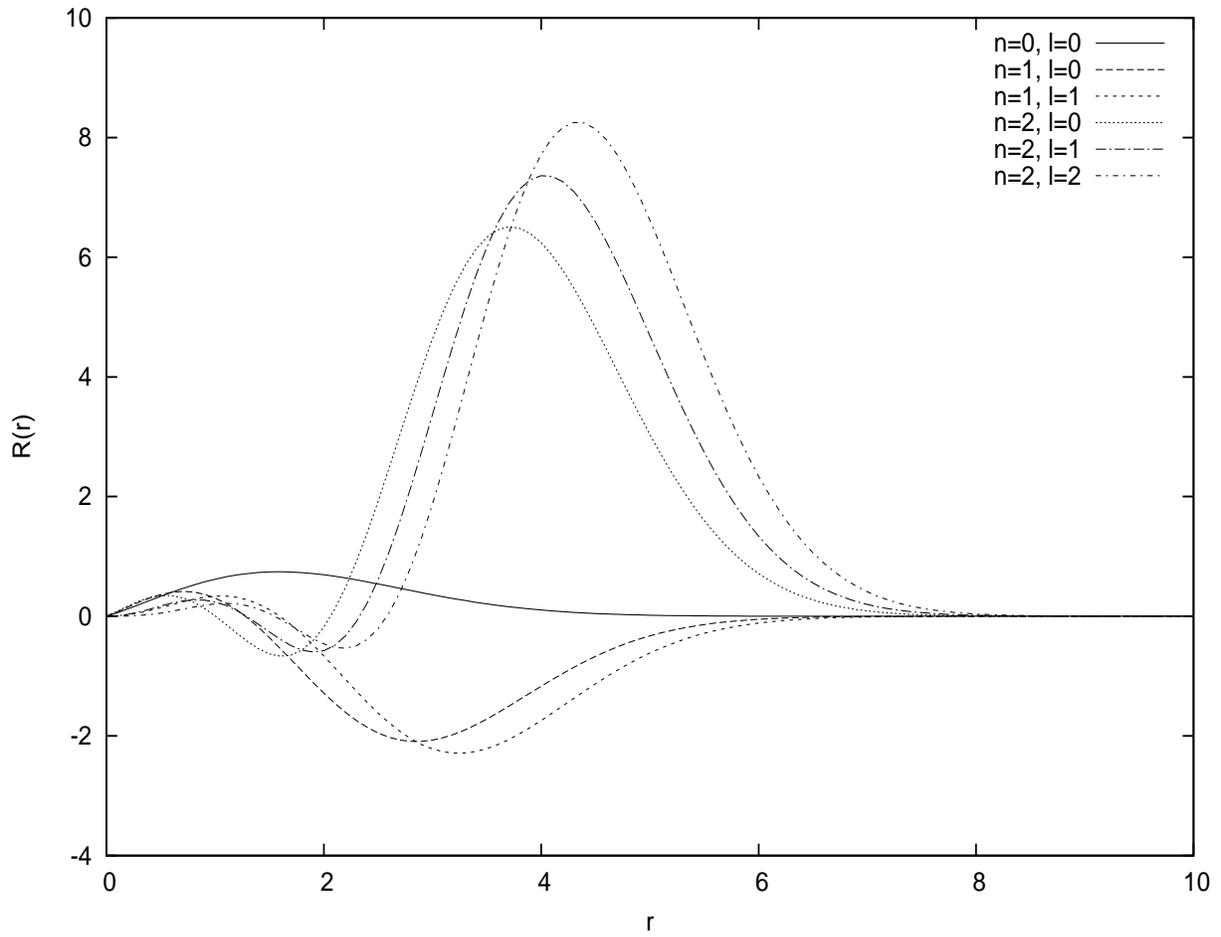}
\caption{Variation of the first six normalized eigenfunctions of
the pseudoharmonic potential (in $m=\hbar=1$ unit).}
\end{figure}

\newpage

\begin{figure}
\centering
\includegraphics[height=5in, width=6.5in, angle=0]{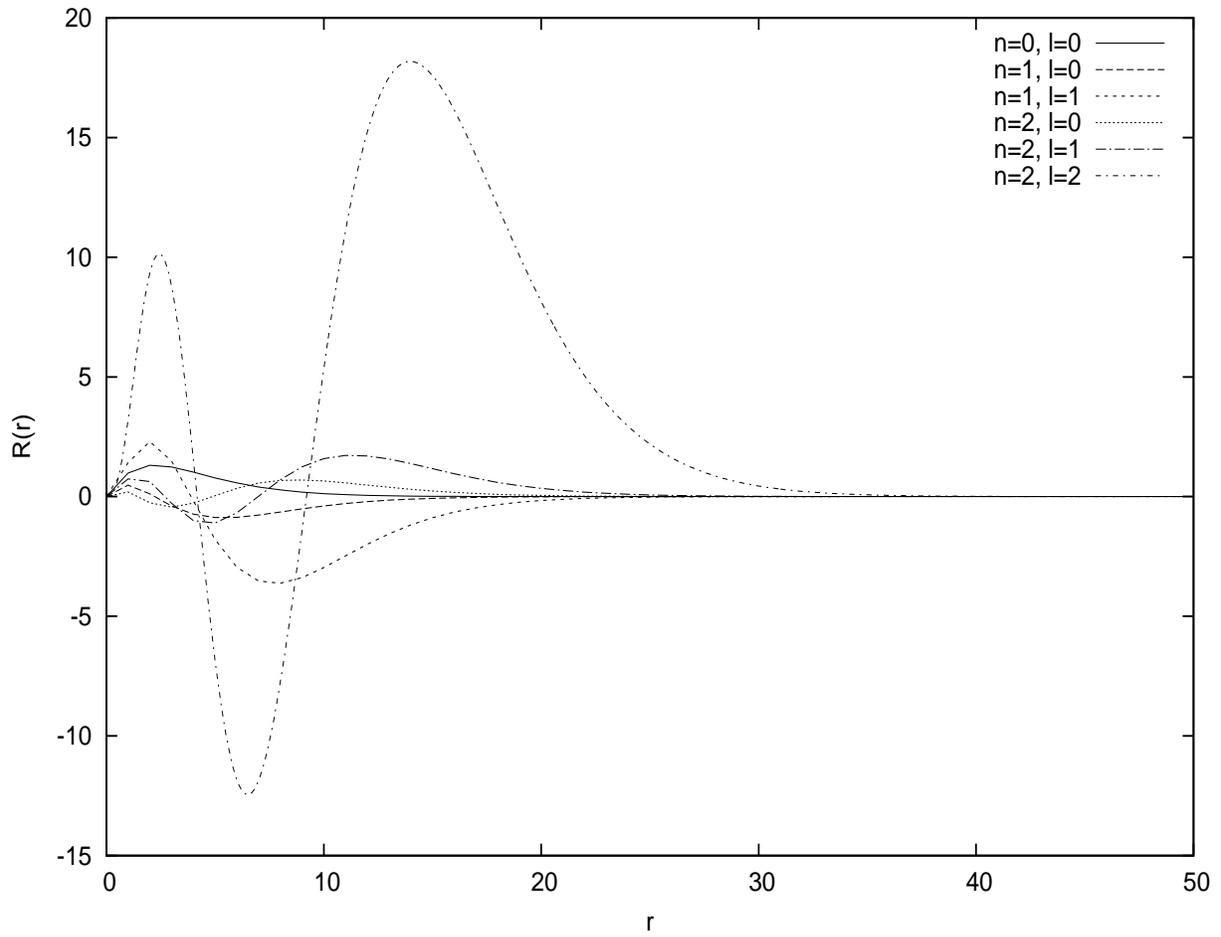}
\caption{Dependence of the first six normalized eigenfunctions of
the Kratzer potential on spatially coordinate $r$ (in $m=\hbar=1$
unit).}
\end{figure}

\end{document}